\documentclass[reprint,superscriptaddress]{revtex4-2}

\usepackage{graphicx}
\usepackage{bm}
\usepackage{amsmath}
\usepackage{amssymb}
\usepackage{braket}
\usepackage[svgnames]{xcolor}

\newcommand{\w}{\omega}
\newcommand{\dif}{\mathrm{d}}

\allowdisplaybreaks

\begin{document}

\title{Feynman integral reduction with intersection theory made simple}

\affiliation{Zhejiang Institute of Modern Physics, School of Physics, Zhejiang University, Hangzhou 310027, China}

\author{Li-Hong Huang}
\email{lhhuang@pku.edu.cn}
\affiliation{School of Physics, Peking University, Beijing 100871, China}

\author{Yan-Qing Ma}
\email{yqma@pku.edu.cn}
\affiliation{School of Physics, Peking University, Beijing 100871, China}

\author{Ziwen Wang}
\email{zwenwang@zju.edu.cn}
\affiliation{Zhejiang Institute of Modern Physics, School of Physics, Zhejiang University, Hangzhou 310027, China}

\author{Li Lin Yang}
\email{yanglilin@zju.edu.cn}
\affiliation{Zhejiang Institute of Modern Physics, School of Physics, Zhejiang University, Hangzhou 310027, China}

\begin{abstract}
Feynman integral reduction based on intersection theory provides an alternative to the traditional integration-by-parts method, yet its practical application has been constrained by the large number of variables required in the computation. In this Letter, we demonstrate that by employing the recently introduced branch representation, the reduction of $L$-loop Feynman integrals with an arbitrary number of external legs can be achieved through the computation of at most $(3L-3)$-variable intersection numbers. This constitutes a significant simplification compared to existing approaches, particularly for multi-leg integrals where the number of variables in conventional methods scales with the total number of propagators. We validate the proposed method through explicit calculations of two-loop diagrams, demonstrating substantial improvements in computational efficiency relative to both traditional intersection-theory approaches and standard integration-by-parts reduction techniques.
\end{abstract}

\maketitle

\section{Introduction}

The integration-by-parts (IBP) reduction~\cite{Tkachov:1981wb, Chetyrkin:1981qh} serves as a foundational tool in the evaluation of Feynman integrals. In contemporary applications, it enables the expression of a vast number of integrals as linear combinations of a significantly smaller set of master integrals (MIs), thereby streamlining complex calculations. The method of differential equations~\cite{Kotikov:1990kg, Kotikov:1991pm, Remiddi:1997ny, Gehrmann:1999as}, a mainstream approach for the analytical evaluation of Feynman integrals, relies crucially on integral reduction as a prerequisite step. Furthermore, IBP reduction constitutes an indispensable component in the numerical evaluation of Feynman integrals through techniques such as the auxiliary mass flow method \cite{Liu:2022chg}.

In the IBP reduction procedure, one must generate and subsequently solve a large system of linear equations, typically employing the Laporta algorithm~\cite{Laporta:2000dc, Laporta:2000dsw}. Several well-developed program packages have been created for this purpose, including \texttt{FIRE}~\cite{Smirnov:2023yhb}, \texttt{Reduze}~\cite{vonManteuffel:2012np}, \texttt{LiteRed}~\cite{Lee:2013mka}, and \texttt{Kira}~\cite{Klappert:2020nbg,Lange:2025fba}. The computational complexity of the linear system escalates considerably as either the number of loops or the number of external legs increases. For cutting-edge problems in high-precision perturbation theory, solving these linear systems demands substantial computational resources and has emerged as a significant bottleneck. Consequently, the development of more efficient approaches for performing IBP reduction remains a pressing priority in the field.

Various strategies have been developed to address this computational challenge. Specialized packages such as \texttt{NeatIBP}~\cite{Wu:2023upw} and \texttt{Blade}~\cite{Guan:2024byi} generate more compact IBP systems by exploiting the algebraic structure inherent in the integrals. Recent investigations have also explored the application of artificial intelligence techniques to optimize integral reduction procedures~\cite{Song:2025pwy}. Despite these notable advances, the reduction of multi-loop multi-leg Feynman integrals continues to present a formidable computational obstacle that necessitates fundamentally novel approaches.

Feynman integral reduction based on intersection theory has been introduced and developed in a series of works~\cite{Mizera:2017rqa, Mastrolia:2018uzb, Frellesvig:2019kgj, Frellesvig:2019uqt, Mizera:2019vvs, Mizera:2020wdt, Frellesvig:2020qot, Caron-Huot:2021xqj, Caron-Huot:2021iev, Chestnov:2022alh, Fontana:2023amt, Brunello:2023rpq, Brunello:2024tqf}. Within this framework, Feynman integrals belonging to a given family are treated as generalized hypergeometric functions taking the form
\begin{equation}
    \int_{\mathcal{C}} u \, \varphi = \int_{\mathcal{C}} u(\bm{z}) \, \hat{\varphi}(\bm{z}) \dif^n \bm{z} \,,
\end{equation}
where $\bm{z}=(z_1,\cdots,z_n)$ denotes the coordinates on the $n$-dimensional base space. The \textit{twist} $u(\bm{z})$ characterizes the integral family and constitutes a multivalued function that vanishes on the boundary of the integration domain $\mathcal{C}$. A specific differential $n$-form $\varphi$ corresponds to a particular Feynman integral within this family. Two $n$-forms are considered equivalent when related by an IBP transformation: $\varphi \sim \varphi + \nabla_\w \xi$, where the covariant derivative is defined as $\nabla_\w \equiv \dif + \omega \wedge$ with the connection $1$-form $\w \equiv \dif \log u$. The equivalence classes of $n$-forms constitute elements of a vector space known as the twisted cohomology group $H_\w^n$. The objective of Feynman integral reduction thus reduces to decomposing a bra-vector $\bra{\varphi}$ (an element of $H_\w^n$) as a linear combination of bra-basis vectors $\{\bra{e_i}\}$ (corresponding to the MIs):
\begin{equation}
    \bra{\varphi} = \sum_{i=1}^{\nu} c_i \bra{e_i} \,,
\end{equation}
where $\nu$ denotes the dimension of $H_\w^n$ and equals the number of master integrals.

To determine the reduction coefficients $\{c_i\}$, one utilizes a dual ket-vector space equipped with a ket-basis $\{\ket{h_j}\}$, and defines the intersection number as a bilinear pairing (inner product) $\braket{\varphi | h_j}_\w$ between a bra-vector $\bra{\varphi}$ and a ket-vector $\ket{h_j}$. The reduction coefficients can then be extracted through the relation
\begin{equation}
    c_i = \braket{\varphi | h_j}_\w \left( \bm{C}^{-1} \right)_{ji} \,,
    \label{eq:reduction_coefficient}
\end{equation}
where $\bm{C}$ represents the metric matrix with elements $C_{ij} \equiv \braket{e_i | h_j}_\w$. The computational efficiency of the reduction procedure depends critically on the complexity of evaluating these $n$-variable intersection numbers.

The computation of $n$-variable intersection numbers requires selecting a fibration of the $n$-dimensional base space, which involves choosing a suitable set of coordinate variables $\bm{z}$ and establishing their computational order. The intersection numbers are then evaluated recursively following this prescribed sequence. Each variable defines a distinct computational layer where specific linear systems must be solved to generate the necessary information for subsequent layers. Consequently, the overall computational complexity depends strongly on the value of $n$, representing the number of variables or layers.

In state-of-the-art applications, Feynman integrals frequently involve $\mathcal{O}(10)$ propagators, resulting in a comparable number of variables $n \sim \mathcal{O}(10)$. This characteristic poses a substantial challenge to the intersection-theory approach for integral reduction. Although the calculational methodology for individual layers has undergone significant improvements over recent years \cite{Fontana:2023amt, Brunello:2023rpq, Brunello:2024tqf}, the large number of layers continues to represent a fundamental bottleneck for this approach.

A recent study~\cite{Lu:2024dsb} has demonstrated that formulating intersection theory within the Feynman parameterization, rather than the Baikov representation, enables a modest reduction in the number of layers by eliminating the need to introduce variables associated with irreducible scalar products. In this Letter, we extend this insight by showing that an appropriate choice of fibration within the Feynman parametrization can effectively reduce the number of layers to $3L-3$ for $L$-loop integrals with an arbitrary number of external legs. This advancement yields a significant simplification for integral reduction utilizing intersection theory.

The remainder of this Letter is organized as follows. In Sec.~\ref{sec:intersection}, we review the essential elements of intersection theory for Feynman integral reduction and introduce the branch representation framework. Section~\ref{sec:dual} presents our construction of dual bases and the methodology for computing their intersection numbers. In Sec.~\ref{sec:examples}, we provide explicit examples demonstrating the practical implementation and computational efficiency of our approach. Finally, we conclude in Sec.~\ref{sec:summary} with a summary and perspectives for future applications.

\section{Intersection numbers in the branch representation}\label{sec:intersection}

Following Ref.~\cite{Lu:2024dsb}, we adopt the Lee-Pomeransky (LP) variant of the Feynman parametrization~\cite{Lee:2013hzt}. In the LP representation, omitting irrelevant prefactors, an $L$-loop Feynman integral can be expressed as
\begin{equation}
    J(\bm{a}) = \int_0^\infty \dif^N \bm{y} \, \mathcal{G}^{-d/2} \prod_i^N y_i^{a_i-1} \,,
    \label{eq:LP_rep}
\end{equation}
where $\bm{y}$ denotes the collection of $N$ Feynman parameters and $\bm{a}$ represents the corresponding propagator powers. The LP polynomial is defined as $\mathcal{G} = \mathcal{U} + \mathcal{F}$, with $\mathcal{U}$ and $\mathcal{F}$ denoting the first and second Symanzik polynomials, respectively. If some $a_i \leq 0$, the integral belongs to a sub-sector living on a relative boundary~\cite{Lu:2024dsb}. As a simple example, $y_i^{-1}$ is regarded as $\lim_{\rho \to 0} \rho \, y_i^{-1+\rho} = \delta(y_i)$.

\begin{figure}[th!]
	\centering
	\includegraphics[width=0.5\linewidth]{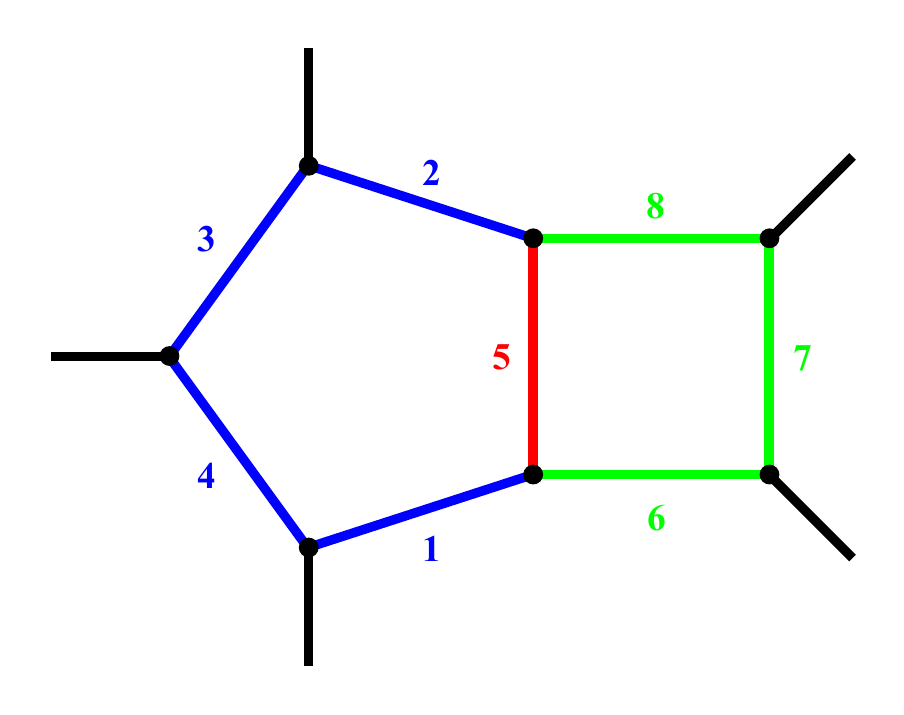}
	\caption{A two-loop diagram illustrating the branch structure. Internal lines sharing the same color belong to the same branch.}
	\label{fig:pentaBOX}
\end{figure}

In an $L$-loop Feynman integral, each propagator contains a quadratic term of the form $\sum_{i,j=1}^L \mathcal{A}_{ij} \, l_i \cdot l_j$, where $l_i$ denote the loop momenta. In Ref.~\cite{Huang:2024nij}, it was observed that this structure can be exploited through a variable transformation, yielding a novel and powerful variant of the Feynman parametrization termed the ``branch representation''. The key insight is that propagators sharing identical quadratic terms belong to the same branch. Figure~\ref{fig:pentaBOX} illustrates this concept for a two-loop diagram, where internal lines of the same color represent propagators belonging to the same branch. The total number of branches $B$ satisfies the bound $B \leq 3L-3$ for $L \geq 2$, a property that proves essential for the efficiency of our method.

For the $b$-th branch, we collect the associated Feynman parameters and denote them $y_{b,\alpha}$. We then introduce the branch variable $X_b = \sum_{\alpha=1}^{N_b} y_{b,\alpha}$, where $N_b$ represents the number of propagators in the $b$-th branch. Through this variable transformation, the integral in Eq.~\eqref{eq:LP_rep} can be recast as
\begin{equation}
    J(\bm{a}) = \int_0^\infty \dif^B \bm{X} \, K(\bm{a}; \bm{X}) \,,
\end{equation}
where $\bm{X} = (X_1,\cdots,X_B)$ denotes the collection of branch variables, and $K(\bm{a};\bm{X})$ represents a ``fixed-branch integral'' (FBI) defined by
\begin{equation}
    K(\bm{a};\bm{X}) \equiv \int \dif^N \bm{y} \, \mathcal{G}^{-d/2} \prod_i^N y_i^{a_i-1} \prod_{b=1}^B \delta \Big( X_b - \sum_{\alpha=1}^{N_b} y_{b,\alpha} \Big) ,
\end{equation} 
which reduces to an $(N-B)$-fold integral upon applying the delta-function constraints.

The crucial observation of Ref.~\cite{Huang:2024nij} is that FBIs exhibit a one-loop-like structure that enables efficient reduction. Specifically, each $K(\bm{a};\bm{X})$ can be expressed as a linear combination of master FBIs:
\begin{equation}
    K(\bm{a};\bm{X}) = \sum_i C_i(\bm{a};\bm{X}) \, F_i(\bm{X}) \,,
    \label{eq:reduction_FBI}
\end{equation}
where the master FBIs $F_i(\bm{X})$ correspond to corner integrals with indices taking values of either $1$ or $0$, and the reduction coefficients $C_i(\bm{a};\bm{X})$ are rational functions of the branch variables $\bm{X}$. While conventional intersection theory would require computing $(N-B)$-variable intersection numbers to obtain these coefficients, the inherent structure of FBIs enables their determination almost ``for free''. Consequently, the complete reduction of the Feynman integral $J(\bm{a})$ requires only the computation of intersection numbers for the branch variables $\bm{X}$. This effectively reduces the problem complexity to computing $B$-variable intersection numbers satisfying the bound $B \leq 3L-3$, irrespective of the number of external legs. We now proceed to describe the technical implementation of this approach.

Within the framework of intersection theory, we identify the master FBIs as constituting the bra-basis for the ``inner layer'', denoted as $\bra{e_i^{(0)}} \equiv \bra{F_i(\bm{X})}$. Without loss of generality, we select $X_1$ as the variable for the first recursive layer, designated as layer-1. The computation of intersection numbers at layer-1, taking the general form $\braket{\varphi | h_j^{(1)}}_{\w^{(1)}}$, requires specification of the ket-basis vectors $\ket{h_j^{(1)}}$ and the connection $1$-form $\w^{(1)}$, with $X_2,\cdots,X_B$ treated as external parameters. The calculation of such intersection numbers depends on three essential ingredients, among which the connection matrix $\Omega^{(1)}$ presents the greatest technical complexity and is therefore discussed first.

The elements of the connection matrix $\Omega^{(1)}$ are given by
\begin{equation}
    \Omega^{(1)}_{ij} = \sum_k \braket{\nabla_{X_1} e_i^{(0)} | h_k^{(0)}}_{\w^{(0)}}  \left( \bm{C}_{(0)}^{-1} \right)_{kj} \,,
    \label{eq:Omega1}
\end{equation}
where $\nabla_{X_1} e_i^{(0)} \equiv \partial_{X_1} e_i^{(0)} + e_i^{(0)} \partial_{X_1} \log u$ with $u=\mathcal{G}^{-d/2}$ representing the twist in the LP representation, $\omega^{(0)}$ denotes the connection $1$-form for the inner layer, and $\ket{h_k^{(0)}}$ are the ket-basis vectors for the inner layer. The metric matrix $\bm{C}_{(0)}$ for the inner layer has elements $\braket{e_i^{(0)} | h_j^{(0)}}_{\w^{(0)}}$. A crucial property is that $\Omega^{(1)}_{ij}$ is independent of the specific choice of ket-basis $\ket{h_k^{(0)}}$, as this dependence cancels between the two factors in Eq.~\eqref{eq:Omega1}. Indeed, $\Omega^{(1)}_{ij}$ can be interpreted as the inner-layer reduction coefficient of $\bra{\nabla_{X_1} e_i^{(0)}}$ projected onto $\bra{e_j^{(0)}}$:
\begin{equation}
    \bra{\nabla_{X_1} e_i^{(0)}} = \sum_j \Omega^{(1)}_{ij} \bra{e_j^{(0)}} \,.
\end{equation}
Since $\bra{e_j^{(0)}}$ corresponds to the FBI $F_j(\bm{X})$, and $\bra{\nabla_{X_1} e_i^{(0)}}$ corresponds to the derivative $\partial_{X_1} F_i(\bm{X})$, the matrix $\Omega^{(1)}$ represents precisely the coefficient matrix appearing in the differential equations satisfied by the master FBIs with respect to $X_1$. This matrix can be obtained efficiently through FBI-reduction as specified in Eq.~\eqref{eq:reduction_FBI}.

The second ingredient required for computing the layer-1 intersection number $\braket{\varphi | h_j^{(1)}}_{\w^{(1)}}$ comprises the projection coefficients of the bra-vector $\bra{\varphi}$ onto the inner basis $\bra{e_i^{(0)}}$, given by
\begin{equation}
    \sum_j \braket{\varphi | h_j^{(0)}}_{\w^{(0)}}  \left( \bm{C}_{(0)}^{-1} \right)_{ji} \,.
\end{equation}
Analogous to the $\Omega^{(1)}$ matrix, these projection coefficients are independent of the choice of ket-basis $\ket{h_j^{(0)}}$ and can be obtained efficiently via FBI-reduction following Eq.~\eqref{eq:reduction_FBI}.

The final ingredient entering $\braket{\varphi | h_j^{(1)}}_{\w^{(1)}}$ is the inner-layer intersection numbers between the inner-layer bra-basis vectors $\bra{e_i^{(0)}}$ and the layer-1 ket-basis vectors $\ket{h_j^{(1)}}$, denoted $\braket{e_i^{(0)} | h_j^{(1)}}_{\w^{(0)}}$. This quantity presents a significant challenge, as it cannot be obtained through FBI-reduction alone, since fixed-branch integrals contain no intrinsic information about the ket-basis structure. At first glance, this appears to present an insurmountable obstacle.

\section{Construction of dual bases and their intersection numbers}\label{sec:dual}

The resolution to this apparent impasse emerges from recognizing that our objective is to determine the reduction coefficients $c_i$ in Eq.~\eqref{eq:reduction_coefficient}, rather than the individual intersection numbers themselves. Importantly, these reduction coefficients are independent of the specific choice of ket-basis vectors, thereby granting us the freedom to construct the ket-basis as deemed appropriate. The key insight of this work is that explicit knowledge of the ket-basis functional forms is unnecessary; it suffices to understand how to perform the reduction.

Building upon this observation, we adopt the assumption that the inner-layer ket-basis is orthonormal to the inner-layer bra-basis (i.e., the master FBIs): $\braket{e_i^{(0)} | h_j^{(0)}}_{\w^{(0)}} = \delta_{ij}$. While we do not require explicit expressions for $h_j^{(0)}$, we can utilize them to construct the ket-basis for layer-1 and subsequent layers. The layer-1 components take the general form $h_j^{(1)} = \sum_k P_{jk}(\bm{X}) \, h_k^{(0)}$ where $P_{jk}(\bm{X})$ are polynomials of the branch variables (exceptional cases requiring additional construction rules will be discussed below). In practice, it is usually enough to take $P_{jk}(\bm{X})$ as monomials. By virtue of the orthonormality condition, we immediately obtain $\braket{e_i^{(0)} | h_j^{(1)}}_{\w^{(0)}} = P_{ji}(\bm{X})$ without requiring knowledge of the explicit forms of $h_k^{(0)}$.

This result completes the necessary ingredients for computing layer-1 intersection numbers. Similar operations can be carried out recursively for subsequent layers involving variables $X_2$ through $X_B$. The intersection numbers at layer-$n$ depend on those computed at layer-$(n-1)$ in preceding steps. Thus, we have demonstrated that the reduction coefficients of $\bra{\varphi}$ can be calculated using $B$-variable intersection numbers. However, one technical subtlety requires further consideration.

This subtlety emerges since ``sub-branch'' integrals appear at layer-1 and beyond, which are absent at the inner layer. This situation relates to an intrinsic property of FBIs: because the branch variable $X_1$ is held fixed as a generic constant in the inner layer, FBIs cannot accommodate integrands containing $\prod_{\alpha=1}^{N_1} \delta(y_{1,\alpha})$. However, at layer-1, where $X_1$ is integrated over, such integrands become permissible. These correspond to integrals where all propagators belonging to the $X_1$-branch are pinched to a point, effectively setting $X_1 = 0$. On the other hand, ket-vectors of the form $\sum_k P_{jk}(\bm{X}) \, h_k^{(0)}$ with $P_{jk}(\bm{X})$ being polynomials cannot correctly pick-up the residue information at $X_1 = 0$, since the inner-layer ket-basis is not allowed to have such singularities.

To address such cases, we need to dive into each sub-branches and construct the bra- and ket-basis vectors separately. At layer-1, we only need to consider the sub-$X_1$ integrals. Suppose that we have a bra-basis vector of the form $e_\star^{(1)} = f_\star(\hat{\bm{y}}) \prod_{\alpha=1}^{N_1} \delta(y_{1,\alpha})$, where $\hat{\bm{y}}$ denotes the variables from other branches, and $f_\star$ is a corner integrand in the sub-branch where $X_1 = 0$. To construct the corresponding ket-basis vector, we note that there exist inner-layer basis vectors in the top-branch of the form
\begin{equation}
    e_i^{(0)} = f_\star(\hat{\bm{y}}) \prod_{\substack{\alpha=1 \\ \alpha \neq i}}^{N_1} \delta(y_{1,\alpha}) \,, \quad (i=1,\cdots,N_1) \,,
\end{equation}
with corresponding orthonormal duals $h_i^{(0)}$. The layer-1 ket-basis vector corresponding to $\bra{e_\star^{(1)}}$ can be taken as
\begin{equation}\label{eq:h-form}
    h_\star^{(1)} = \frac{1}{X_1} \sum_{i=1}^{N_1} h_i^{(0)} \,.
\end{equation}
This construction yields $\braket{e_\star^{(1)} | h_\star^{(1)}}_{\w^{(1)}} = 1$, and the intersection numbers with other bra-vectors can be evaluated using the methods outlined above. This construction scheme can be applied recursively to subsequent layers.

\section{Examples}\label{sec:examples}

\begin{figure}[th!]
	\centering
	\includegraphics[width=0.5\linewidth]{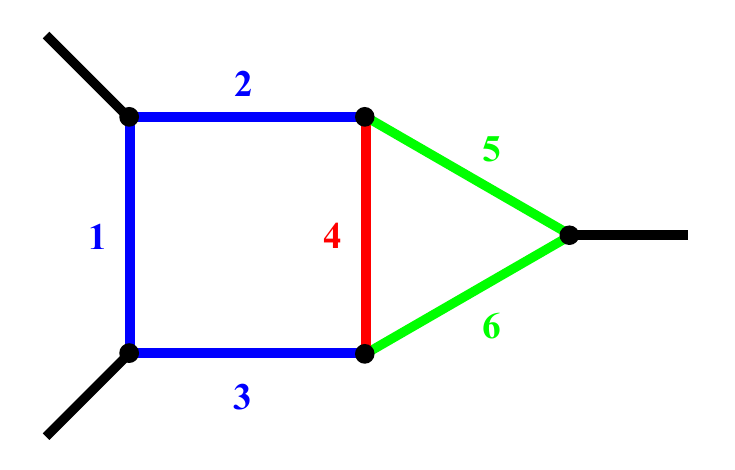}
	\caption{A two-loop three-point diagram with massive internal lines and off-shell external legs.}
	\label{fig:triangle}
\end{figure}

To compare computational efficiency across different representations within the intersection-theory framework, we first consider the three-point diagram shown in Fig.~\ref{fig:triangle}, with off-shell external legs and massive internal lines, where all propagators have the same mass except for the fourth one. In the LP representation, this example requires six layers of intersection numbers. In contrast, our branch representation reduces the number of layers to the theoretical minimum $3L-3 = 3$ for any two-loop diagram, avoiding the exponential growth of complexity associated with additional layers. For the computation of single-layer intersection numbers, we have implemented a proof-of-concept program with \texttt{FiniteFlow} \cite{Peraro:2019svx}, adopting the state-of-the-art method based on polynomial divisions and companion matrices \cite{Fontana:2023amt, Brunello:2023rpq, Brunello:2024tqf}. To deal with analytic regulators, we adopt the approach described in Sec.~4 of Ref.~\cite{Fontana:2023amt}, where a regulator is introduced only for the variable of the current layer and the limit $\rho\to 0$ is taken immediately at that layer. In this way, the $\rho$-dependence does not enter the finite field reconstruction \footnote{In principle, we may also adopt the relative-cohomology based approach described in \cite{Lu:2024dsb} for the LP representation. However, in that case one needs to compute intersection numbers separately for each sub-sector (relative boundary). The large number of sub-sectors in multi-leg integral families makes the implementation rather cumbersome, which also weakens the potential gain in efficiency.}.

The above implementation produces correct reduction coefficients, in both the LP and the branch representations. In the LP representation, the runtime is 10785 seconds (with kinematic invariants taken as rational numbers) on a desktop computer with a 12-core AMD Ryzen 9 5900X CPU. By contrast, our branch-representation based method leads to a runtime of 285 seconds on the same machine, achieving an improvement in efficiency by a factor of 38. This speedup for a simple topology already demonstrates the power of our approach. It can be expected that in a more complicated topology with more propagators, the speedup must be more pronounced.

To assess the potential of our method in cutting-edge problems, we consider the two-loop pentabox diagram in Fig.~\ref{fig:pentaBOX}. All internal masses are set to zero, while the five external legs are taken to be off-shell. For this diagram, intersection number computations need 11 layers in the Baikov representation and 8 layers in the LP representation. Such computations (with our in-house program) are well beyond the computing resources available to us. In contrast, the branch-representation based method reduces the number of layers to the theoretical minimum of $3L-3 = 3$. In this case, the inner-layer dimension of the twisted cohomology group spanned by the FBIs is 105. We choose the order of the branch variables as
\begin{equation}
	X_1 = y_1 + y_2 + y_3 + y_4 \,, \; X_2 = y_5 \,, \; X_3 = y_6 + y_7 +y_8 \,,
\end{equation}
and the dimensions of these three layers are 210, 445 and 228, respectively. After imposing symmetry relations, the total number of master integrals is reduced from 228 to 216. These symmetries can be taken into account after the computation of intersection numbers. With this setup, our program manages to obtain the correct reduction coefficients within a reasonable amount of time.

While it's not possible to directly compare the runtime among different intersection-theory based approaches for complicated integral families, it is worth investigating the potential of our method compared to traditional IBP methods. In essence, the computation of layer-$n$ intersection numbers amounts to the solution of a set of linear systems constructed according to the poles of the connection matrix $\Omega^{(n)}$. As is well-known, the computational complexity for solving a system of $N$ linear equations scales as $\mathcal{O}(N^3)$. It is therefore meaningful to compare the sizes of the linear systems generated in the computation of intersection numbers with those generated by IBP programs such as \texttt{Kira~3}~\cite{Lange:2025fba}.

We consider the pentabox family in Fig.~\ref{fig:pentaBOX}, with reduction targets with no numerators and up to 2 dots, e.g., $J(1, 1, 1, 1, 1, 2, 2, 1)$. In \texttt{Kira 3}, for these targets, we need to set \{\texttt{r:\,10, s:\,1, d:\,2}\}, and the program generates a linear system with approximately $1.9 \times 10^5$ equations. On the other hand, our intersection-number computations need to solve a set of much smaller linear systems, with one system with the size $1.1 \times 10^4$, about 100 systems with sizes of $\mathcal{O}(10^3)$, and a couple of smaller ones. To make a comparison, we define an effective size $N_{\text{eff}}$ through $N_{\text{eff}}^3=\sum_{i} N_i^3$, where $N_i$ is the size of the $i$-th system. For the current problem, we find $N_{\mathrm{eff}} \sim 1.3\times 10^4$, which serves as a rough measure of the overall complexity of intersection-number computations. We note that this is more than an order of magnitude smaller than the size of the linear system generated by \texttt{Kira 3}. This demonstrate the great potential of the intersection-number based method built upon the branch representation for Feynman integral reduction. Furthermore, the linear systems generated in intersection-number computations are automatically in a block triangular and sparse form -- in stark contrast to the unstructured systems generated by traditional IBP methods. Employing these properties, the computational burden can be further brought down to the level of $\sum_{i} N_i^2$. We anticipate that it is promising to build an optimized implementation based on our method to be competitive with or even surpass the current industrial standards for Feynman integral reduction.

\section{Summary and outlook}\label{sec:summary}

In this Letter, we have proposed a novel method for Feynman integral reduction based on intersection theory. By utilizing the recently introduced branch representation, we have demonstrated that the reduction of $L$-loop Feynman integrals effectively reduces to the computation of $(3L-3)$-variable intersection numbers, independent of the number of external legs. This constitutes a significant theoretical advance, as the number of variables in conventional approaches scales with the total number of propagators. Our explicit calculations for two-loop diagrams demonstrate that this reduction in computational complexity translates into substantial practical improvements in efficiency compared to both traditional intersection-theory methods and standard integration-by-parts reduction. The method proves particularly promising for multi-leg integrals relevant to multi-boson and multi-jet production processes at the Large Hadron Collider and future high-energy colliders. Future work will extend this approach to higher-loop integrals and develop optimized numerical implementations for large-scale phenomenological applications.

\begin{acknowledgments}
We thank Hjalte Frellesvig and Xuhang Jiang for valuable discussions. This work was supported in part by the National Natural Science Foundation of China under Grant No. 12325503, 12375097, 12535003, 12547104, and the Fundamental Research Funds for the Central Universities.
\end{acknowledgments}

\bibliographystyle{apsrev4-1}
\bibliography{references_inspire}

\begin{thebibliography}{35}%
\makeatletter
\providecommand \@ifxundefined [1]{%
 \@ifx{#1\undefined}
}%
\providecommand \@ifnum [1]{%
 \ifnum #1\expandafter \@firstoftwo
 \else \expandafter \@secondoftwo
 \fi
}%
\providecommand \@ifx [1]{%
 \ifx #1\expandafter \@firstoftwo
 \else \expandafter \@secondoftwo
 \fi
}%
\providecommand \natexlab [1]{#1}%
\providecommand \enquote  [1]{``#1''}%
\providecommand \bibnamefont  [1]{#1}%
\providecommand \bibfnamefont [1]{#1}%
\providecommand \citenamefont [1]{#1}%
\providecommand \href@noop [0]{\@secondoftwo}%
\providecommand \href [0]{\begingroup \@sanitize@url \@href}%
\providecommand \@href[1]{\@@startlink{#1}\@@href}%
\providecommand \@@href[1]{\endgroup#1\@@endlink}%
\providecommand \@sanitize@url [0]{\catcode `\\12\catcode `\$12\catcode
  `\&12\catcode `\#12\catcode `\^12\catcode `\_12\catcode `\%12\relax}%
\providecommand \@@startlink[1]{}%
\providecommand \@@endlink[0]{}%
\providecommand \url  [0]{\begingroup\@sanitize@url \@url }%
\providecommand \@url [1]{\endgroup\@href {#1}{\urlprefix }}%
\providecommand \urlprefix  [0]{URL }%
\providecommand \Eprint [0]{\href }%
\providecommand \doibase [0]{http://dx.doi.org/}%
\providecommand \selectlanguage [0]{\@gobble}%
\providecommand \bibinfo  [0]{\@secondoftwo}%
\providecommand \bibfield  [0]{\@secondoftwo}%
\providecommand \translation [1]{[#1]}%
\providecommand \BibitemOpen [0]{}%
\providecommand \bibitemStop [0]{}%
\providecommand \bibitemNoStop [0]{.\EOS\space}%
\providecommand \EOS [0]{\spacefactor3000\relax}%
\providecommand \BibitemShut  [1]{\csname bibitem#1\endcsname}%
\let\auto@bib@innerbib\@empty
\bibitem [{\citenamefont {Tkachov}(1981)}]{Tkachov:1981wb}%
  \BibitemOpen
  \bibfield  {author} {\bibinfo {author} {\bibfnamefont {F.~V.}\ \bibnamefont
  {Tkachov}},\ }\href {\doibase 10.1016/0370-2693(81)90288-4} {\bibfield
  {journal} {\bibinfo  {journal} {Phys. Lett. B}\ }\textbf {\bibinfo {volume}
  {100}},\ \bibinfo {pages} {65} (\bibinfo {year} {1981})}\BibitemShut
  {NoStop}%
\bibitem [{\citenamefont {Chetyrkin}\ and\ \citenamefont
  {Tkachov}(1981)}]{Chetyrkin:1981qh}%
  \BibitemOpen
  \bibfield  {author} {\bibinfo {author} {\bibfnamefont {K.~G.}\ \bibnamefont
  {Chetyrkin}}\ and\ \bibinfo {author} {\bibfnamefont {F.~V.}\ \bibnamefont
  {Tkachov}},\ }\href {\doibase 10.1016/0550-3213(81)90199-1} {\bibfield
  {journal} {\bibinfo  {journal} {Nucl. Phys. B}\ }\textbf {\bibinfo {volume}
  {192}},\ \bibinfo {pages} {159} (\bibinfo {year} {1981})}\BibitemShut
  {NoStop}%
\bibitem [{\citenamefont {Kotikov}(1991{\natexlab{a}})}]{Kotikov:1990kg}%
  \BibitemOpen
  \bibfield  {author} {\bibinfo {author} {\bibfnamefont {A.~V.}\ \bibnamefont
  {Kotikov}},\ }\href {\doibase 10.1016/0370-2693(91)90413-K} {\bibfield
  {journal} {\bibinfo  {journal} {Phys. Lett. B}\ }\textbf {\bibinfo {volume}
  {254}},\ \bibinfo {pages} {158} (\bibinfo {year}
  {1991}{\natexlab{a}})}\BibitemShut {NoStop}%
\bibitem [{\citenamefont {Kotikov}(1991{\natexlab{b}})}]{Kotikov:1991pm}%
  \BibitemOpen
  \bibfield  {author} {\bibinfo {author} {\bibfnamefont {A.~V.}\ \bibnamefont
  {Kotikov}},\ }\href {\doibase 10.1016/0370-2693(91)90536-Y} {\bibfield
  {journal} {\bibinfo  {journal} {Phys. Lett. B}\ }\textbf {\bibinfo {volume}
  {267}},\ \bibinfo {pages} {123} (\bibinfo {year} {1991}{\natexlab{b}})},\
  \bibinfo {note} {[Erratum: Phys.Lett.B 295, 409--409 (1992)]}\BibitemShut
  {NoStop}%
\bibitem [{\citenamefont {Remiddi}(1997)}]{Remiddi:1997ny}%
  \BibitemOpen
  \bibfield  {author} {\bibinfo {author} {\bibfnamefont {E.}~\bibnamefont
  {Remiddi}},\ }\href {\doibase 10.1007/BF03185566} {\bibfield  {journal}
  {\bibinfo  {journal} {Nuovo Cim. A}\ }\textbf {\bibinfo {volume} {110}},\
  \bibinfo {pages} {1435} (\bibinfo {year} {1997})},\ \Eprint
  {http://arxiv.org/abs/hep-th/9711188} {arXiv:hep-th/9711188} \BibitemShut
  {NoStop}%
\bibitem [{\citenamefont {Gehrmann}\ and\ \citenamefont
  {Remiddi}(2000)}]{Gehrmann:1999as}%
  \BibitemOpen
  \bibfield  {author} {\bibinfo {author} {\bibfnamefont {T.}~\bibnamefont
  {Gehrmann}}\ and\ \bibinfo {author} {\bibfnamefont {E.}~\bibnamefont
  {Remiddi}},\ }\href {\doibase 10.1016/S0550-3213(00)00223-6} {\bibfield
  {journal} {\bibinfo  {journal} {Nucl. Phys. B}\ }\textbf {\bibinfo {volume}
  {580}},\ \bibinfo {pages} {485} (\bibinfo {year} {2000})},\ \Eprint
  {http://arxiv.org/abs/hep-ph/9912329} {arXiv:hep-ph/9912329} \BibitemShut
  {NoStop}%
\bibitem [{\citenamefont {Liu}\ and\ \citenamefont {Ma}(2023)}]{Liu:2022chg}%
  \BibitemOpen
  \bibfield  {author} {\bibinfo {author} {\bibfnamefont {X.}~\bibnamefont
  {Liu}}\ and\ \bibinfo {author} {\bibfnamefont {Y.-Q.}\ \bibnamefont {Ma}},\
  }\href {\doibase 10.1016/j.cpc.2022.108565} {\bibfield  {journal} {\bibinfo
  {journal} {Comput. Phys. Commun.}\ }\textbf {\bibinfo {volume} {283}},\
  \bibinfo {pages} {108565} (\bibinfo {year} {2023})},\ \Eprint
  {http://arxiv.org/abs/2201.11669} {arXiv:2201.11669 [hep-ph]} \BibitemShut
  {NoStop}%
\bibitem [{\citenamefont {Laporta}(2001)}]{Laporta:2000dc}%
  \BibitemOpen
  \bibfield  {author} {\bibinfo {author} {\bibfnamefont {S.}~\bibnamefont
  {Laporta}},\ }\href {\doibase 10.1016/S0370-2693(01)00256-8} {\bibfield
  {journal} {\bibinfo  {journal} {Phys. Lett. B}\ }\textbf {\bibinfo {volume}
  {504}},\ \bibinfo {pages} {188} (\bibinfo {year} {2001})},\ \Eprint
  {http://arxiv.org/abs/hep-ph/0102032} {arXiv:hep-ph/0102032} \BibitemShut
  {NoStop}%
\bibitem [{\citenamefont {Laporta}(2000)}]{Laporta:2000dsw}%
  \BibitemOpen
  \bibfield  {author} {\bibinfo {author} {\bibfnamefont {S.}~\bibnamefont
  {Laporta}},\ }\href {\doibase 10.1142/S0217751X00002159} {\bibfield
  {journal} {\bibinfo  {journal} {Int. J. Mod. Phys. A}\ }\textbf {\bibinfo
  {volume} {15}},\ \bibinfo {pages} {5087} (\bibinfo {year} {2000})},\ \Eprint
  {http://arxiv.org/abs/hep-ph/0102033} {arXiv:hep-ph/0102033} \BibitemShut
  {NoStop}%
\bibitem [{\citenamefont {Smirnov}\ and\ \citenamefont
  {Zeng}(2024)}]{Smirnov:2023yhb}%
  \BibitemOpen
  \bibfield  {author} {\bibinfo {author} {\bibfnamefont {A.~V.}\ \bibnamefont
  {Smirnov}}\ and\ \bibinfo {author} {\bibfnamefont {M.}~\bibnamefont {Zeng}},\
  }\href {\doibase 10.1016/j.cpc.2024.109261} {\bibfield  {journal} {\bibinfo
  {journal} {Comput. Phys. Commun.}\ }\textbf {\bibinfo {volume} {302}},\
  \bibinfo {pages} {109261} (\bibinfo {year} {2024})},\ \Eprint
  {http://arxiv.org/abs/2311.02370} {arXiv:2311.02370 [hep-ph]} \BibitemShut
  {NoStop}%
\bibitem [{\citenamefont {von Manteuffel}\ and\ \citenamefont
  {Studerus}(2012)}]{vonManteuffel:2012np}%
  \BibitemOpen
  \bibfield  {author} {\bibinfo {author} {\bibfnamefont {A.}~\bibnamefont {von
  Manteuffel}}\ and\ \bibinfo {author} {\bibfnamefont {C.}~\bibnamefont
  {Studerus}},\ }\href@noop {} {\  (\bibinfo {year} {2012})},\ \Eprint
  {http://arxiv.org/abs/1201.4330} {arXiv:1201.4330 [hep-ph]} \BibitemShut
  {NoStop}%
\bibitem [{\citenamefont {Lee}(2014)}]{Lee:2013mka}%
  \BibitemOpen
  \bibfield  {author} {\bibinfo {author} {\bibfnamefont {R.~N.}\ \bibnamefont
  {Lee}},\ }\href {\doibase 10.1088/1742-6596/523/1/012059} {\bibfield
  {journal} {\bibinfo  {journal} {J. Phys. Conf. Ser.}\ }\textbf {\bibinfo
  {volume} {523}},\ \bibinfo {pages} {012059} (\bibinfo {year} {2014})},\
  \Eprint {http://arxiv.org/abs/1310.1145} {arXiv:1310.1145 [hep-ph]}
  \BibitemShut {NoStop}%
\bibitem [{\citenamefont {Klappert}\ \emph {et~al.}(2021)\citenamefont
  {Klappert}, \citenamefont {Lange}, \citenamefont {Maierh\"ofer},\ and\
  \citenamefont {Usovitsch}}]{Klappert:2020nbg}%
  \BibitemOpen
  \bibfield  {author} {\bibinfo {author} {\bibfnamefont {J.}~\bibnamefont
  {Klappert}}, \bibinfo {author} {\bibfnamefont {F.}~\bibnamefont {Lange}},
  \bibinfo {author} {\bibfnamefont {P.}~\bibnamefont {Maierh\"ofer}}, \ and\
  \bibinfo {author} {\bibfnamefont {J.}~\bibnamefont {Usovitsch}},\ }\href
  {\doibase 10.1016/j.cpc.2021.108024} {\bibfield  {journal} {\bibinfo
  {journal} {Comput. Phys. Commun.}\ }\textbf {\bibinfo {volume} {266}},\
  \bibinfo {pages} {108024} (\bibinfo {year} {2021})},\ \Eprint
  {http://arxiv.org/abs/2008.06494} {arXiv:2008.06494 [hep-ph]} \BibitemShut
  {NoStop}%
\bibitem [{\citenamefont {Lange}\ \emph {et~al.}(2025)\citenamefont {Lange},
  \citenamefont {Usovitsch},\ and\ \citenamefont {Wu}}]{Lange:2025fba}%
  \BibitemOpen
  \bibfield  {author} {\bibinfo {author} {\bibfnamefont {F.}~\bibnamefont
  {Lange}}, \bibinfo {author} {\bibfnamefont {J.}~\bibnamefont {Usovitsch}}, \
  and\ \bibinfo {author} {\bibfnamefont {Z.}~\bibnamefont {Wu}},\ }\href@noop
  {} {\  (\bibinfo {year} {2025})},\ \Eprint {http://arxiv.org/abs/2505.20197}
  {arXiv:2505.20197 [hep-ph]} \BibitemShut {NoStop}%
\bibitem [{\citenamefont {Wu}\ \emph {et~al.}(2024)\citenamefont {Wu},
  \citenamefont {Boehm}, \citenamefont {Ma}, \citenamefont {Xu},\ and\
  \citenamefont {Zhang}}]{Wu:2023upw}%
  \BibitemOpen
  \bibfield  {author} {\bibinfo {author} {\bibfnamefont {Z.}~\bibnamefont
  {Wu}}, \bibinfo {author} {\bibfnamefont {J.}~\bibnamefont {Boehm}}, \bibinfo
  {author} {\bibfnamefont {R.}~\bibnamefont {Ma}}, \bibinfo {author}
  {\bibfnamefont {H.}~\bibnamefont {Xu}}, \ and\ \bibinfo {author}
  {\bibfnamefont {Y.}~\bibnamefont {Zhang}},\ }\href {\doibase
  10.1016/j.cpc.2023.108999} {\bibfield  {journal} {\bibinfo  {journal}
  {Comput. Phys. Commun.}\ }\textbf {\bibinfo {volume} {295}},\ \bibinfo
  {pages} {108999} (\bibinfo {year} {2024})},\ \Eprint
  {http://arxiv.org/abs/2305.08783} {arXiv:2305.08783 [hep-ph]} \BibitemShut
  {NoStop}%
\bibitem [{\citenamefont {Guan}\ \emph {et~al.}(2024)\citenamefont {Guan},
  \citenamefont {Liu}, \citenamefont {Ma},\ and\ \citenamefont
  {Wu}}]{Guan:2024byi}%
  \BibitemOpen
  \bibfield  {author} {\bibinfo {author} {\bibfnamefont {X.}~\bibnamefont
  {Guan}}, \bibinfo {author} {\bibfnamefont {X.}~\bibnamefont {Liu}}, \bibinfo
  {author} {\bibfnamefont {Y.-Q.}\ \bibnamefont {Ma}}, \ and\ \bibinfo {author}
  {\bibfnamefont {W.-H.}\ \bibnamefont {Wu}},\ }\href@noop {} {\  (\bibinfo
  {year} {2024})},\ \Eprint {http://arxiv.org/abs/2405.14621} {arXiv:2405.14621
  [hep-ph]} \BibitemShut {NoStop}%
\bibitem [{\citenamefont {Song}\ \emph {et~al.}(2025)\citenamefont {Song},
  \citenamefont {Yang}, \citenamefont {Cao}, \citenamefont {Luo},\ and\
  \citenamefont {Zhu}}]{Song:2025pwy}%
  \BibitemOpen
  \bibfield  {author} {\bibinfo {author} {\bibfnamefont {Z.-Y.}\ \bibnamefont
  {Song}}, \bibinfo {author} {\bibfnamefont {T.-Z.}\ \bibnamefont {Yang}},
  \bibinfo {author} {\bibfnamefont {Q.-H.}\ \bibnamefont {Cao}}, \bibinfo
  {author} {\bibfnamefont {M.-x.}\ \bibnamefont {Luo}}, \ and\ \bibinfo
  {author} {\bibfnamefont {H.~X.}\ \bibnamefont {Zhu}},\ }\href@noop {} {\
  (\bibinfo {year} {2025})},\ \Eprint {http://arxiv.org/abs/2502.09544}
  {arXiv:2502.09544 [hep-ph]} \BibitemShut {NoStop}%
\bibitem [{\citenamefont {Mizera}(2018)}]{Mizera:2017rqa}%
  \BibitemOpen
  \bibfield  {author} {\bibinfo {author} {\bibfnamefont {S.}~\bibnamefont
  {Mizera}},\ }\href {\doibase 10.1103/PhysRevLett.120.141602} {\bibfield
  {journal} {\bibinfo  {journal} {Phys. Rev. Lett.}\ }\textbf {\bibinfo
  {volume} {120}},\ \bibinfo {pages} {141602} (\bibinfo {year} {2018})},\
  \Eprint {http://arxiv.org/abs/1711.00469} {arXiv:1711.00469 [hep-th]}
  \BibitemShut {NoStop}%
\bibitem [{\citenamefont {Mastrolia}\ and\ \citenamefont
  {Mizera}(2019)}]{Mastrolia:2018uzb}%
  \BibitemOpen
  \bibfield  {author} {\bibinfo {author} {\bibfnamefont {P.}~\bibnamefont
  {Mastrolia}}\ and\ \bibinfo {author} {\bibfnamefont {S.}~\bibnamefont
  {Mizera}},\ }\href {\doibase 10.1007/JHEP02(2019)139} {\bibfield  {journal}
  {\bibinfo  {journal} {JHEP}\ }\textbf {\bibinfo {volume} {02}},\ \bibinfo
  {pages} {139} (\bibinfo {year} {2019})},\ \Eprint
  {http://arxiv.org/abs/1810.03818} {arXiv:1810.03818 [hep-th]} \BibitemShut
  {NoStop}%
\bibitem [{\citenamefont {Frellesvig}\ \emph
  {et~al.}(2019{\natexlab{a}})\citenamefont {Frellesvig}, \citenamefont
  {Gasparotto}, \citenamefont {Laporta}, \citenamefont {Mandal}, \citenamefont
  {Mastrolia}, \citenamefont {Mattiazzi},\ and\ \citenamefont
  {Mizera}}]{Frellesvig:2019kgj}%
  \BibitemOpen
  \bibfield  {author} {\bibinfo {author} {\bibfnamefont {H.}~\bibnamefont
  {Frellesvig}}, \bibinfo {author} {\bibfnamefont {F.}~\bibnamefont
  {Gasparotto}}, \bibinfo {author} {\bibfnamefont {S.}~\bibnamefont {Laporta}},
  \bibinfo {author} {\bibfnamefont {M.~K.}\ \bibnamefont {Mandal}}, \bibinfo
  {author} {\bibfnamefont {P.}~\bibnamefont {Mastrolia}}, \bibinfo {author}
  {\bibfnamefont {L.}~\bibnamefont {Mattiazzi}}, \ and\ \bibinfo {author}
  {\bibfnamefont {S.}~\bibnamefont {Mizera}},\ }\href {\doibase
  10.1007/JHEP05(2019)153} {\bibfield  {journal} {\bibinfo  {journal} {JHEP}\
  }\textbf {\bibinfo {volume} {05}},\ \bibinfo {pages} {153} (\bibinfo {year}
  {2019}{\natexlab{a}})},\ \Eprint {http://arxiv.org/abs/1901.11510}
  {arXiv:1901.11510 [hep-ph]} \BibitemShut {NoStop}%
\bibitem [{\citenamefont {Frellesvig}\ \emph
  {et~al.}(2019{\natexlab{b}})\citenamefont {Frellesvig}, \citenamefont
  {Gasparotto}, \citenamefont {Mandal}, \citenamefont {Mastrolia},
  \citenamefont {Mattiazzi},\ and\ \citenamefont
  {Mizera}}]{Frellesvig:2019uqt}%
  \BibitemOpen
  \bibfield  {author} {\bibinfo {author} {\bibfnamefont {H.}~\bibnamefont
  {Frellesvig}}, \bibinfo {author} {\bibfnamefont {F.}~\bibnamefont
  {Gasparotto}}, \bibinfo {author} {\bibfnamefont {M.~K.}\ \bibnamefont
  {Mandal}}, \bibinfo {author} {\bibfnamefont {P.}~\bibnamefont {Mastrolia}},
  \bibinfo {author} {\bibfnamefont {L.}~\bibnamefont {Mattiazzi}}, \ and\
  \bibinfo {author} {\bibfnamefont {S.}~\bibnamefont {Mizera}},\ }\href
  {\doibase 10.1103/PhysRevLett.123.201602} {\bibfield  {journal} {\bibinfo
  {journal} {Phys. Rev. Lett.}\ }\textbf {\bibinfo {volume} {123}},\ \bibinfo
  {pages} {201602} (\bibinfo {year} {2019}{\natexlab{b}})},\ \Eprint
  {http://arxiv.org/abs/1907.02000} {arXiv:1907.02000 [hep-th]} \BibitemShut
  {NoStop}%
\bibitem [{\citenamefont {Mizera}\ and\ \citenamefont
  {Pokraka}(2020)}]{Mizera:2019vvs}%
  \BibitemOpen
  \bibfield  {author} {\bibinfo {author} {\bibfnamefont {S.}~\bibnamefont
  {Mizera}}\ and\ \bibinfo {author} {\bibfnamefont {A.}~\bibnamefont
  {Pokraka}},\ }\href {\doibase 10.1007/JHEP02(2020)159} {\bibfield  {journal}
  {\bibinfo  {journal} {JHEP}\ }\textbf {\bibinfo {volume} {02}},\ \bibinfo
  {pages} {159} (\bibinfo {year} {2020})},\ \Eprint
  {http://arxiv.org/abs/1910.11852} {arXiv:1910.11852 [hep-th]} \BibitemShut
  {NoStop}%
\bibitem [{\citenamefont {Mizera}(2019)}]{Mizera:2020wdt}%
  \BibitemOpen
  \bibfield  {author} {\bibinfo {author} {\bibfnamefont {S.}~\bibnamefont
  {Mizera}},\ }\href {\doibase 10.22323/1.383.0016} {\bibfield  {journal}
  {\bibinfo  {journal} {PoS}\ }\textbf {\bibinfo {volume} {MA2019}},\ \bibinfo
  {pages} {016} (\bibinfo {year} {2019})},\ \Eprint
  {http://arxiv.org/abs/2002.10476} {arXiv:2002.10476 [hep-th]} \BibitemShut
  {NoStop}%
\bibitem [{\citenamefont {Frellesvig}\ \emph {et~al.}(2021)\citenamefont
  {Frellesvig}, \citenamefont {Gasparotto}, \citenamefont {Laporta},
  \citenamefont {Mandal}, \citenamefont {Mastrolia}, \citenamefont
  {Mattiazzi},\ and\ \citenamefont {Mizera}}]{Frellesvig:2020qot}%
  \BibitemOpen
  \bibfield  {author} {\bibinfo {author} {\bibfnamefont {H.}~\bibnamefont
  {Frellesvig}}, \bibinfo {author} {\bibfnamefont {F.}~\bibnamefont
  {Gasparotto}}, \bibinfo {author} {\bibfnamefont {S.}~\bibnamefont {Laporta}},
  \bibinfo {author} {\bibfnamefont {M.~K.}\ \bibnamefont {Mandal}}, \bibinfo
  {author} {\bibfnamefont {P.}~\bibnamefont {Mastrolia}}, \bibinfo {author}
  {\bibfnamefont {L.}~\bibnamefont {Mattiazzi}}, \ and\ \bibinfo {author}
  {\bibfnamefont {S.}~\bibnamefont {Mizera}},\ }\href {\doibase
  10.1007/JHEP03(2021)027} {\bibfield  {journal} {\bibinfo  {journal} {JHEP}\
  }\textbf {\bibinfo {volume} {03}},\ \bibinfo {pages} {027} (\bibinfo {year}
  {2021})},\ \Eprint {http://arxiv.org/abs/2008.04823} {arXiv:2008.04823
  [hep-th]} \BibitemShut {NoStop}%
\bibitem [{\citenamefont {Caron-Huot}\ and\ \citenamefont
  {Pokraka}(2021)}]{Caron-Huot:2021xqj}%
  \BibitemOpen
  \bibfield  {author} {\bibinfo {author} {\bibfnamefont {S.}~\bibnamefont
  {Caron-Huot}}\ and\ \bibinfo {author} {\bibfnamefont {A.}~\bibnamefont
  {Pokraka}},\ }\href {\doibase 10.1007/JHEP12(2021)045} {\bibfield  {journal}
  {\bibinfo  {journal} {JHEP}\ }\textbf {\bibinfo {volume} {12}},\ \bibinfo
  {pages} {045} (\bibinfo {year} {2021})},\ \Eprint
  {http://arxiv.org/abs/2104.06898} {arXiv:2104.06898 [hep-th]} \BibitemShut
  {NoStop}%
\bibitem [{\citenamefont {Caron-Huot}\ and\ \citenamefont
  {Pokraka}(2022)}]{Caron-Huot:2021iev}%
  \BibitemOpen
  \bibfield  {author} {\bibinfo {author} {\bibfnamefont {S.}~\bibnamefont
  {Caron-Huot}}\ and\ \bibinfo {author} {\bibfnamefont {A.}~\bibnamefont
  {Pokraka}},\ }\href {\doibase 10.1007/JHEP04(2022)078} {\bibfield  {journal}
  {\bibinfo  {journal} {JHEP}\ }\textbf {\bibinfo {volume} {04}},\ \bibinfo
  {pages} {078} (\bibinfo {year} {2022})},\ \Eprint
  {http://arxiv.org/abs/2112.00055} {arXiv:2112.00055 [hep-th]} \BibitemShut
  {NoStop}%
\bibitem [{\citenamefont {Chestnov}\ \emph {et~al.}(2022)\citenamefont
  {Chestnov}, \citenamefont {Gasparotto}, \citenamefont {Mandal}, \citenamefont
  {Mastrolia}, \citenamefont {Matsubara-Heo}, \citenamefont {Munch},\ and\
  \citenamefont {Takayama}}]{Chestnov:2022alh}%
  \BibitemOpen
  \bibfield  {author} {\bibinfo {author} {\bibfnamefont {V.}~\bibnamefont
  {Chestnov}}, \bibinfo {author} {\bibfnamefont {F.}~\bibnamefont
  {Gasparotto}}, \bibinfo {author} {\bibfnamefont {M.~K.}\ \bibnamefont
  {Mandal}}, \bibinfo {author} {\bibfnamefont {P.}~\bibnamefont {Mastrolia}},
  \bibinfo {author} {\bibfnamefont {S.~J.}\ \bibnamefont {Matsubara-Heo}},
  \bibinfo {author} {\bibfnamefont {H.~J.}\ \bibnamefont {Munch}}, \ and\
  \bibinfo {author} {\bibfnamefont {N.}~\bibnamefont {Takayama}},\ }\href
  {\doibase 10.1007/JHEP09(2022)187} {\bibfield  {journal} {\bibinfo  {journal}
  {JHEP}\ }\textbf {\bibinfo {volume} {09}},\ \bibinfo {pages} {187} (\bibinfo
  {year} {2022})},\ \Eprint {http://arxiv.org/abs/2204.12983} {arXiv:2204.12983
  [hep-th]} \BibitemShut {NoStop}%
\bibitem [{\citenamefont {Fontana}\ and\ \citenamefont
  {Peraro}(2023)}]{Fontana:2023amt}%
  \BibitemOpen
  \bibfield  {author} {\bibinfo {author} {\bibfnamefont {G.}~\bibnamefont
  {Fontana}}\ and\ \bibinfo {author} {\bibfnamefont {T.}~\bibnamefont
  {Peraro}},\ }\href {\doibase 10.1007/JHEP08(2023)175} {\bibfield  {journal}
  {\bibinfo  {journal} {JHEP}\ }\textbf {\bibinfo {volume} {08}},\ \bibinfo
  {pages} {175} (\bibinfo {year} {2023})},\ \Eprint
  {http://arxiv.org/abs/2304.14336} {arXiv:2304.14336 [hep-ph]} \BibitemShut
  {NoStop}%
\bibitem [{\citenamefont {Brunello}\ \emph
  {et~al.}(2024{\natexlab{a}})\citenamefont {Brunello}, \citenamefont
  {Chestnov}, \citenamefont {Crisanti}, \citenamefont {Frellesvig},
  \citenamefont {Mandal},\ and\ \citenamefont {Mastrolia}}]{Brunello:2023rpq}%
  \BibitemOpen
  \bibfield  {author} {\bibinfo {author} {\bibfnamefont {G.}~\bibnamefont
  {Brunello}}, \bibinfo {author} {\bibfnamefont {V.}~\bibnamefont {Chestnov}},
  \bibinfo {author} {\bibfnamefont {G.}~\bibnamefont {Crisanti}}, \bibinfo
  {author} {\bibfnamefont {H.}~\bibnamefont {Frellesvig}}, \bibinfo {author}
  {\bibfnamefont {M.~K.}\ \bibnamefont {Mandal}}, \ and\ \bibinfo {author}
  {\bibfnamefont {P.}~\bibnamefont {Mastrolia}},\ }\href {\doibase
  10.1007/JHEP09(2024)015} {\bibfield  {journal} {\bibinfo  {journal} {JHEP}\
  }\textbf {\bibinfo {volume} {09}},\ \bibinfo {pages} {015} (\bibinfo {year}
  {2024}{\natexlab{a}})},\ \Eprint {http://arxiv.org/abs/2401.01897}
  {arXiv:2401.01897 [hep-th]} \BibitemShut {NoStop}%
\bibitem [{\citenamefont {Brunello}\ \emph
  {et~al.}(2024{\natexlab{b}})\citenamefont {Brunello}, \citenamefont
  {Chestnov},\ and\ \citenamefont {Mastrolia}}]{Brunello:2024tqf}%
  \BibitemOpen
  \bibfield  {author} {\bibinfo {author} {\bibfnamefont {G.}~\bibnamefont
  {Brunello}}, \bibinfo {author} {\bibfnamefont {V.}~\bibnamefont {Chestnov}},
  \ and\ \bibinfo {author} {\bibfnamefont {P.}~\bibnamefont {Mastrolia}},\
  }\href@noop {} {\  (\bibinfo {year} {2024}{\natexlab{b}})},\ \Eprint
  {http://arxiv.org/abs/2408.16668} {arXiv:2408.16668 [hep-th]} \BibitemShut
  {NoStop}%
\bibitem [{\citenamefont {Lu}\ \emph {et~al.}(2025)\citenamefont {Lu},
  \citenamefont {Wang},\ and\ \citenamefont {Yang}}]{Lu:2024dsb}%
  \BibitemOpen
  \bibfield  {author} {\bibinfo {author} {\bibfnamefont {M.}~\bibnamefont
  {Lu}}, \bibinfo {author} {\bibfnamefont {Z.}~\bibnamefont {Wang}}, \ and\
  \bibinfo {author} {\bibfnamefont {L.~L.}\ \bibnamefont {Yang}},\ }\href
  {\doibase 10.1007/JHEP05(2025)158} {\bibfield  {journal} {\bibinfo  {journal}
  {JHEP}\ }\textbf {\bibinfo {volume} {05}},\ \bibinfo {pages} {158} (\bibinfo
  {year} {2025})},\ \Eprint {http://arxiv.org/abs/2411.05226} {arXiv:2411.05226
  [hep-th]} \BibitemShut {NoStop}%
\bibitem [{\citenamefont {Lee}\ and\ \citenamefont
  {Pomeransky}(2013)}]{Lee:2013hzt}%
  \BibitemOpen
  \bibfield  {author} {\bibinfo {author} {\bibfnamefont {R.~N.}\ \bibnamefont
  {Lee}}\ and\ \bibinfo {author} {\bibfnamefont {A.~A.}\ \bibnamefont
  {Pomeransky}},\ }\href {\doibase 10.1007/JHEP11(2013)165} {\bibfield
  {journal} {\bibinfo  {journal} {JHEP}\ }\textbf {\bibinfo {volume} {11}},\
  \bibinfo {pages} {165} (\bibinfo {year} {2013})},\ \Eprint
  {http://arxiv.org/abs/1308.6676} {arXiv:1308.6676 [hep-ph]} \BibitemShut
  {NoStop}%
\bibitem [{\citenamefont {Huang}\ \emph {et~al.}(2024)\citenamefont {Huang},
  \citenamefont {Huang},\ and\ \citenamefont {Ma}}]{Huang:2024nij}%
  \BibitemOpen
  \bibfield  {author} {\bibinfo {author} {\bibfnamefont {L.-H.}\ \bibnamefont
  {Huang}}, \bibinfo {author} {\bibfnamefont {R.-J.}\ \bibnamefont {Huang}}, \
  and\ \bibinfo {author} {\bibfnamefont {Y.-Q.}\ \bibnamefont {Ma}},\
  }\href@noop {} {\  (\bibinfo {year} {2024})},\ \Eprint
  {http://arxiv.org/abs/2412.21053} {arXiv:2412.21053 [hep-ph]} \BibitemShut
  {NoStop}%
\bibitem [{\citenamefont {Peraro}(2019)}]{Peraro:2019svx}%
  \BibitemOpen
  \bibfield  {author} {\bibinfo {author} {\bibfnamefont {T.}~\bibnamefont
  {Peraro}},\ }\href {\doibase 10.1007/JHEP07(2019)031} {\bibfield  {journal}
  {\bibinfo  {journal} {JHEP}\ }\textbf {\bibinfo {volume} {07}},\ \bibinfo
  {pages} {031} (\bibinfo {year} {2019})},\ \Eprint
  {http://arxiv.org/abs/1905.08019} {arXiv:1905.08019 [hep-ph]} \BibitemShut
  {NoStop}%
\bibitem [{Note1()}]{Note1}%
  \BibitemOpen
  \bibinfo {note} {In principle, we may also adopt the relative-cohomology
  based approach described in \cite {Lu:2024dsb} for the LP representation.
  However, in that case one needs to compute intersection numbers separately
  for each sub-sector (relative boundary). The large number of sub-sectors in
  multi-leg integral families makes the implementation rather cumbersome, which
  also weakens the potential gain in efficiency.}\BibitemShut {Stop}%
\end{thebibliography}%

\end{document}